\documentclass[preprint]{aastex}
\usepackage{lscape}
\begin{document}

\title{Searching for Planets in the Hyades V: Limits on Planet Detection in the 
Presence of Stellar Activity
\footnote{Data presented herein were obtained at the 
W.M. Keck Observatory, which is operated as a scientific partnership among
the California Institute of Technology, the University of California and 
the National Aeronautics and Space Administration. The Observatory was made
possible by the generous financial support of the W.M. Keck Foundation.
Additional data were obtained with the Hobby-Eberly Telescope which is 
operated by McDonald Observatory on behalf of the University of Texas at Austin,
the Pennsylvania State University, Stanford University, Ludwig-Maximilians-
Universit\"at M\"unchen, and Georg-August-Universit\"at G\"ottingen.}}

\author{Diane B. Paulson\altaffilmark{2}}
\affil{Department of Astronomy, University of Texas, Austin, TX 78712}
\email{apodis@umich.edu}
\altaffiltext{2}{Current Address: Department of Astronomy, University of 
Michigan, Ann Arbor, MI 48109}

\author{William D. Cochran}
\affil{McDonald Observatory, University of Texas, Austin, TX 78712}
\email{wdc@astro.as.utexas.edu}

\and
\author{Artie P. Hatzes}
\affil{Th\"{u}ringer Landessternwarte Tautenburg, D-07778 Tautenburg, 
Germany}

\begin{abstract}
We present the results of a radial velocity survey of a sample of Hyades stars,
and discuss the effects of stellar activity on radial velocity measurements. 
The level of radial velocity scatter due to rotational modulation of 
stellar surface features for the Hyades is in agreement with the 
predictions of \citet{SaDo97}- the maximum radial velocity rms
of up to $\sim$50~m~s$^{-1}$, with an average rms of 
$\sim$16~m~s$^{-1}$. In this sample of 94 stars, we find 1 new 
binary, 2 stars with linear trends indicative of binary companions, and no 
close-in giant planets. We discuss the limits on extrasolar planet detection 
in the Hyades and the constraints imposed on radial velocity surveys of 
young stars. 
\end{abstract}


\section{Introduction}

Radial velocity ($v_{\rm r}$) surveys for extrasolar planets have been
extremely successful (e.g. \nocite{BuMaWi96} Butler et al. 1996).
These surveys have, however, largely
excluded young, active stars \citep{VoBuMa02, CuMaBu99, SaDo97}.
The reason given was that the activity levels of
young stars is significant enough to cause large variations in the measured
$v_{\rm r}$. Although it does not introduce a true $v_{\rm r}$ shift
\citep[e.g. Saar \& Donahue 1997, hereafter SD97,][]{Ha02}, the apparent 
shift is caused by a change in the line shape of the absorption features.
SD97 quantify the predicted amplitude of this phenomenon, and it 
has been observationally confirmed by several groups 
\citep[e.g.][]{QuHeSi01,HeDoBa02, PaSaCo04, SaFi00, SaBuMa98}.
While detection of extrasolar planets around young stars will be 
complicated by these spectral line profile variations, there is much to 
learn about the frequency of planets and their 
orbital characteristics at all stellar ages. So, it is necessary to 
learn the limitations of the techniques employed in planet detection and 
then proceed (if possible) with planet searches.

We present here the results of the radial velocity search for extrasolar planets
in a sample of Hyades dwarfs. Primarily, we discuss the mean level of 
radial velocity noise caused by stellar magnetic activity and the 
possibilities of detecting planets in Hyades-aged stars. 
 
\section{Observations and Analysis}
We have been studying a sample of Hyades
dwarfs ranging from spectral classes F5 to M2 with the Keck I High Resolution
Echelle Spectrometer \citep{VoAlBi94}. 
The observations and analysis of the $v_{\rm r}$ data are discussed
in \citet{CoHaPa02}.
While we made every attempt to
include only stars which were not binaries, we discovered some
stars which have too high $v_{\rm r}$ rms to be non-binaries. These stars are
discussed in \S 5.2.
 
The measurement of $v_{\rm r}$ involves using an
I$_{2}$ gas absorption cell as a standard velocity reference \citep{VaBuMa95}.
A signal-to-noise ratio
(S/N) of $\sim$150-300 is achieved at 5000\AA\ with
resolution $R\simeq$60,000. In the case of high S/N ($\ge$200), we achieve an 
internal
precision $\sim$3--4~m~s$^{-1}$ for a given star \citep{CoHaPa02}; while a 
spectrum with S/N$\sim$100 yields precision of $\sim$6~m~s$^{-1}$. In
addition, the exposure times are maintained at $\leq$15 minutes.
We use standard IRAF
packages to reduce the CCD images and extract the observed
spectra. The $v_{\rm r}$ measurements are made using a program called
RADIAL (developed at the University of
Texas, UT, and McDonald Observatory) to measure precise
radial velocities. This program was adapted for use with data
from all of the planet search programs affiliated with UT. Discussions
of RADIAL may be found in \citet{CoHaBu97} and \citet{HaCoMA00}. The $v_{\rm r}$
measurements of all data in this sample obtained with the Keck telescope 
are listed in Table 1. Measurements of vB~15, vB~18 and vB~153 taken with the 
HRS at the Hobby-Eberly Telescope (HET) are discussed and listed in 
\nocite{PaSaCo04}Paulson et al. (2004) (hereafter PSC04).

\section{Results}
While the target sample was selected in part on non-binarity, a
handful of binaries did end up in our sample.
The stars listed in Table 2 show either significant linear trends (most
likely binaries) or a defined binary orbit (vB~88).  The measured $v_{\rm r}$
of binary stars and stars with
significant linear trends are shown in Figure 1.  Three
binaries discovered by \citet{PaGhRe98} were not removed from the
sample in the inital compilation of targets. These are noted in the
Column 3 of Table 2. Stars which only have a small slope
(i.e. those with $\sigma_{v_{\rm r}}\sim40$~m~s$^{-1}$) have
not been included in this table.
vB~88 appears to have almost completed 1 orbit. A {\itshape tenative}
solution (shown in Figure 1) gives an $m$sin$i$=0.07~M$_{\odot}$. 
Using the measured $v$sin$i$ for vB~88 and 
and an estimate of the true rotational velocity as derived by \citet{PaSnCo03},
we estimate the mass of the companion to be $\sim$0.86~$\pm$~0.31~M$_{\odot}$,
most likely a K dwarf, to the F8V parent star. While 0.31~M$_{\odot}$ is
the formal error, we note that if the mass were much larger than 
0.86~M$_{\odot}$, we
would see a double lined spectrum and we see no indication of this
in the vB~88 spectra. So, 0.86~M$_{\odot}$ is 
most likely an upper limit to the true mass.
The orbital parameters 
for this binary companion are listed in Table 3. 
 
With three exceptions discussed later in this section (those showing 
significant
long-period trends) and 1 star with poor sampling but with velocity rms of
72~m~s$^{-1}$, the remaining stars show no significant linear trends
(with rms ($\sigma_{v_{\rm r}}$)$\le$40~m~s$^{-1}$).
Table 4 lists the program stars, the rms of the observations
with internal errors removed ($\sigma_{v_{\rm r, int}}$) and
the average internal error of the observations for each
star ($\sigma_{int}$). This is a summary of the 
observations presented in Table 1. Figure 2 shows a histogram of the
$\sigma_{v_{\rm r}}$ for the program stars excluding binaries and
stars with linear trends. The internal $v_{\rm r}$ errors have 
been removed. We note that the majority of stars have
$\sim 5 \le \sigma_{v_{\rm r}} \le$ 25 m~s$^{-1}$, which is what we
expect from stars of this activity level and age (SD97, PSC04).
In PSC04, we showed that some stars in
the sample display $v_{\rm r}$ variations of $\sim$40~m~s$^{-1}$ due
to stellar active regions. Therefore, the stars with $\sigma_{v_{\rm r}} \sim
50~\rm m~s^{-1}$ could also suffer from severe effects of activity. 
It is of interest that we do not find any stars in this sample with very large
$\sigma_{v_{\rm r}}$ ($\ge$100 m~s$^{-1}$) with suggestive short periods, thus
no ``hot Jupiters" with mass $\ge$1 M$_{Jup}$.
 
To explore the spread in $\sigma_{v_{\rm r}}$ in the sample, we compared
the $\sigma_{v_{\rm r}}$ of each star with the measured rotational velocity
($v$sin$i$) from \citet{PaSnCo03}.
Although a little less than half of the program stars
have measured $v$sin$i$, we are still able to see an obvious trend in the
data (see Figure 3- $\sigma_{v_{\rm r}}$ versus $v$sin$i$, with binaries
and stars with linear trends excluded and internal $v_{\rm r}$ errors
removed).  While there is significant scatter in 
the figure, a trend is still present. 
The location of active regions, the fraction
of the surface covered by activity, the rotational velocity and the
inclination of the stars all play significant roles in
the measured $v_{\rm r}$. The higher the $v$sin$i$, the more broadened the
spectral absorption features will be. The internal error becomes
larger with increased $v$sin$i$ (see Figure 4), as determination of the
line center becomes increasingly difficult. As the inclination of the star
decreases (becomes pole-on) lower $\sigma_{v_{\rm r}}$ is expected. This is
because the most significant effect of active regions in this age of
star (e.g. \nocite{SaDo97}SD97) is short-period 
variations due to the
rotational modulation of the features across the stellar surface. Take a
simple case which assumes active regions are equatorial.
When the star is pole-on (0$^{\circ}$), variations of this nature will
diminish and similarly, when the star is face-on (90$^{\circ}$), the effect
will be at a maximum. Certainly, the larger fraction of the stellar surface
covered by active regions will cause larger amplitude variations and the
physical location of the features will cause variable effects on the
line shape (and thus, the measured $v_{\rm r}$). It is not surprising, then,
that there are a few stars with very large $\sigma_{v_{\rm r}}$, as
statistically, a few stars should have very low inclinations. This effect of
increased $\sigma_{v_{\rm r}}$ with increased $v$sin$i$ is also predicted
by A. Hatzes (2003, private communication) and discussed in SD97.
Additionally, scatter in $\sigma_{v_{\rm r}}$ (in Figure 3) at a given
$v$sin$i$ is caused by star-to-star variations in the overall activity level.
Although the stars do not show cyclic behavior in chromospheric activity,
the overall level of activity varies.
 
As also shown in PSC04, some Hyades stars (vB~153, in particular) may
show long lived active regions or active longitudes. Therefore, should we
detect significant periods compatible with an expected  or observed rotation
rate, we could explore the nature of stars which display this type of
activity.
Therefore, to the remaining stars, we employed the same period-finding
algorithm as used in PSC04 
(that of \nocite{HoBa86}Horne \& Baliunas 1986). From this technique we are able to determine the
most significant periods in a data set and the false alarm probabilities (FAP)
associated with each period. We also perform a bootstrap algorithm (e.g.
\nocite{KuScCu97}K{\"u}rster et al. 1997)
to determine a false alarm probability based solely on the
data. The bootstrap method does not assume a gaussian noise distribution
as do the more traditional FAPs (such as Horne \& Baliunas 1986).
This method randomizes the data-
keeping  the observed times (in JD) the same but randomly assigning the
$v_{\rm r}$ observed to those JDs. The resultant ``fake'' data are run through
the periodogram to determine the most significant period of this data.
This randomization process is iterated 1000 times and a false alarm probability
is the ratio of the number of times the power of the detected period in
the fake data was equal to or larger than the power of the signal in the
original data. If the data are pure noise, then the false alarm probability
should be very high ($\sim$1.0). The results of this analysis are listed in
Table 5 for stars which had FAPs of less than 10\% in each of the 
two methods (Horne \& Baliunas and bootstrap). We explore the
actual significance of these periods in $\S$~4.
 
It is not surprising that the periods we derive here are not consistent
with the published values of rotation periods ($P_{\rm rot}$)
because active regions evolve.
Table
6 lists the periods we derive versus those from literature (references
included in the table notes) and predicted values from \citet{DuFrLa84}, where
possible.
Additionlly, the periods derived from the Keck data for the stars which we
have dervied $P_{\rm rot}$ (from Paulson et al. 2004 for
vB~15, vB~18 and vB~153)
are inconsistent with each other (also see Table 6).
Thus, significant phase variations have occured in the span of 
observations of the data taken from Keck.
Because we do not have sufficient temporal coverage, we are unable to
perform detailed studies of the evolution of these
active regions- the decay time or migration of the active regions.
 
\section{Limits on Substellar Companions}
Using the methods outlined in \citet[ hereafter NA98]{NeAn98} we compute
analytical limits on
the detectability of substellar mass companions based on the duration and
accuracy of the data for each star. We then compare this to a
periodogram-based analytical derivation to find significant peaks representing
the significant signals in the data. At frequencies
where peaks cross the analytical limits, the detection of periodic signals
(either companions or periodic activity) is present at the 99\% confidence
level.
 
From Eq. 4 of NA98, the companion mass is $M_{\rm C}$~sin$i$=$K(P
M_{\star}^{2}(2\pi G)^{-1})^{1/3}$, where $i$ is the inclination of the
companion, $P$ is the period sampled, $K$ is the velocity amplitude, 
$M_{\star}$ is the mass of the parent star, and $G$ is the
gravitational constant.
We derive $K$ and $P$ from the above periodogram analysis and $M_{\star}$
is estimated from B-V measurements \citep{APLa99} and \citet{Gr92}.
We will refer to the resultant $M_{\rm C}$~sin$i$ as the
``companion mass power spectrum'' (solid curve in Figure 5).
 
To define the limits on companion masses detected (or significant
periodicity from intrinsic sources), we must calculate the
limit on the detectable velocity.
$K_{\rm X}$ (Eq. 15 in NA98) is the velocity amplitude exceeded by any
of the $N$ fits to randomized data in a given period range with
probability 1$-$$X$, where $X$ is the product of probabilities of each fit
at each period sampled ($X~=~\cal P^{\rm n}$) and $n$ is the number of
sampled periods. Thus, $K_{\rm X} = 2\sigma(n^{-1}~\rm ln\ (2\pi P_{0}
(f_{\rm 1}~-~f_{\rm 2})\rm (1~-~X)^{-1}))^{1/2}$, where $P_{0}$ is
the duration of the original data (JD$_{\rm final} - \rm JD_{\rm begin}$) and
$f_{\rm 1}$ and $f_{\rm 2}$ are the lower and upper limits, respectively,
of the period intervals (in our case, 4 bins per dex). $X$ is determined from
the bootstrap ``false alarm probability'' described in $\S$~3.
Using $K_{X}$ now in place of $K$ in Eq. 4 (NA98), we obtain the 99\%
confidence level of the mass limits of companions. Shown in Figure 5 are
representative cases of these calculations. We show both the Keck and
HET data with two different error assumptions, described below, for 3 stars.
The Keck data for all stars show the same behavior as the three stars shown in 
Figure 5.  Thus, we only show here these three because we also have
data from the HET for each. The histogram plot limits have been suppressed at 
1 year and 11 years for HET and Keck data, respectively.
 
The dashed histogram lines in Figure 5 assume that the only error
in the observations are internal effects ($\sim$4-7~m~s$^{-1}$). In the HET
data (left column), we discover significant peaks crossing the 99\%
threshold. This tells us that there are significant periods at these
crossings. However, these significant periods do not
correspond to companions but to the rotational period of the star and
aliases thereof. For the Keck data (right column), there are several crossings.
This is expected as the sampling is extremely poor. There are $\sim$15-20
observations of these stars over the course of 6 years, and many ``significant''
periods can be derived with this quality of sampling.
                                                                                
A far superior assumption, and that recommended by NA98,
 is that the error in the observations is not
only due to internal errors but also due to the rotational modulation of
active regions. For this, we assume an error equal to the mean $v_{\rm r}$
rms for the sample ($\sim$16~m~s$^{-1}$, ignoring stars with linear
trends and binary stars). Thus, any period
spikes that cross the 99\% threshold {\itshape should} come from external
sources (stellar and substellar companions). The calculation using this
assumption is depicted by the solid histogram lines in Figure 5. We note that
for all non-binary stars, there are absolutely
no companion mass power spectrum crossings of the 99\% threshold for 
stars in these Keck data. The binary
stars have insufficent phase
coverage. So, the threshold crossings are beyond the reasonable
limits of the calculations described above.
 
Thus, we can use this method
for determining the value of ``significant'' periods derived from
a periodogram analysis quite apart from a calculated false alarm probability.

In general, we can provide constraints on the characteristics of systems 
which are detectable around young stars via the radial velocity technique, also
employing the methods in NA98. Figure 6 presents an analysis for stars of
Hyades age with different sampling. Both panels show the $K$ velocity 
semi-amplitude versus orbital period. The histograms are the limits for 
a 99\% confidence detection. The solid histogram (a) represents sampling similar
to that which we observed from Keck, while the dotted histogram (b) is similar
to the sampling we obtained from HET. The dotted histogram has been 
suppressed beyond 1 year. The two curves shown are for reference-
the solid curve (c) is a 1 M$_{Jup}$ companion and the dotted curve (d) is a
3 M$_{Jup}$ companion. These curves are the $K$ velocities of companions given
that they are in orbits with zero eccentricity  and are companions to a 1 
M$_{\odot}$ star with 90$^{\circ}$ inclination. The velocity-period space
laying above the histograms are detectable with 99\% confidence. So, for 
example, the top panel representing data with errors (radial velocity jitter
from activity + internal errors) of 
$\sigma$=16~m~s$^{-1}$ shows that a 1~M$_{Jup}$ companion
is only detectable if it has periods less than $\sim$100 days with only a few
observations a year for 5 years and is detectable 
if with longer orbital periods only if the sampling is quite good 
(several observations 
a month). On the other hand, the bottom panel shows a more realistic case, 
as the true noise from activity will be higher than 16~m~s$^{-1}$ 
if the system is edge on
(corresponding to the $K$ curves c and d which are for edge on orbits). For 
data with errors of $\sigma$=40~m~s$^{-1}$, the detectabiliy of planets with 
$\lesssim$1 M$_{Jup}$ becomes impossible for poor sampling. For young stars
with such high levels of activity related radial velocity noise,
it is only feasible to look for either very high mass 
companions or the data must be taken with extremely good sampling (several 
observations a month for $\sim$3 months).

\section{Discussion}
We can determine significant period in data by various techniques, including
those discussed in this paper. But, it is most useful to understand when
significant periods are real or simply artifacts of sampling. The analysis
of the periodogram produces periods with FAPs~$\sim$10\% for several stars.
Phasing the data to these periods produces periodic curves (by-eye inspection).
This is inadequate. Therefore, we have employed the method of \citet{NeAn98}
to explore the significance of detections. All
short-periods detected turn out to be artifacts of the sampling and of the
quality of the data. 
 
The detection of planets around young stars is complicated by the rotational
modulation of 
stellar active regions. The activity not only causes high levels of $v_{\rm r}$
noise but
can also yield periodic variations in the measured $v_{\rm r}$ causing
false detections. The procedure we adopted \citep{NeAn98} picks out all
significant signals given the quantity and quality of data,
so we must be careful in the identification of the
source of variability. 
In our data, we find no evidence for short-period massive planets 
or brown dwarfs.  Finally, of the 94 stars in this
sample, 6 are either suspected or identified binaries and 1 has a velocity
rms which is somewhat arge but further observations are required to say 
anything more concrete- it is still within possible ``jitter" from
high activity levels. 

Future detection of 
extrasolar planets around young stars via the radial velocity method will 
be limited to 
high-mass planets and in particular, those with short orbital periods. 
Constraints on telescope time needed for these surveys becomes clear. In order 
to increase the odds of planet detection, as current planet searches
have determined that only $\sim$1\% of stars do have ``hot Jupiters",
data must be sampled several times a month which requires a great deal of 
allocated telescope time.

\acknowledgments
This research was partially supported by a grant from NASA administered by
the American Astronomical Society. DBP and WDC are also supported by NASA grant
NAG5-9227 and NSF grant AST-9808980. We would like to thank the anonymous 
referee for very useful suggestions in preparing this manuscript for 
publication.


\begin{figure}
\plotone{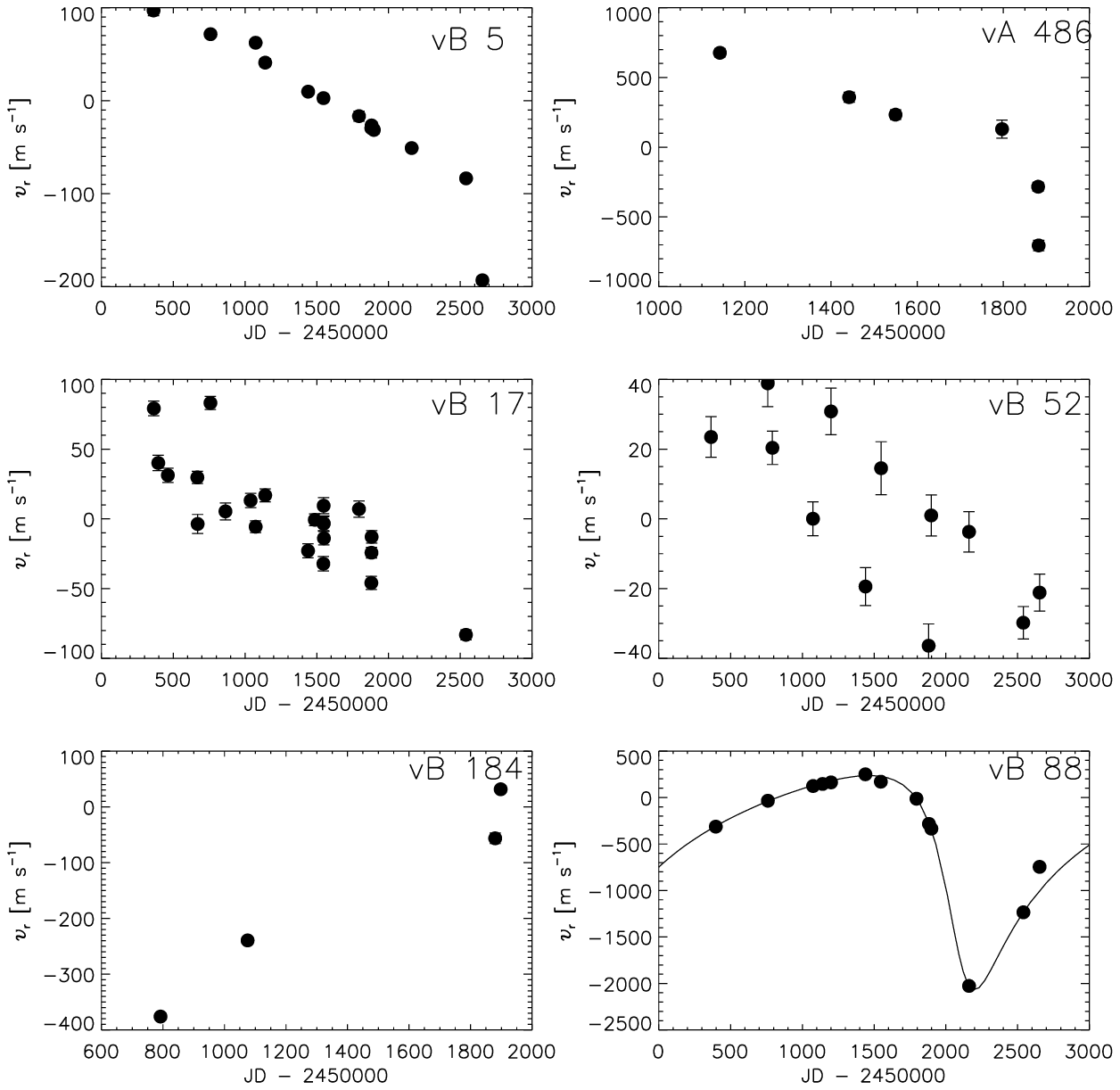}
\caption[Known Binaries and Stars with Linear Trends]{Known binaries and
stars with linear trends (suspected binaries).
The orbital fit to the $v_{\rm r}$ data for vB 88 is shown along with its
$v_{\rm r}$ data. Internal error bars are shown.}
\end{figure}
 
\begin{figure}
\plotone{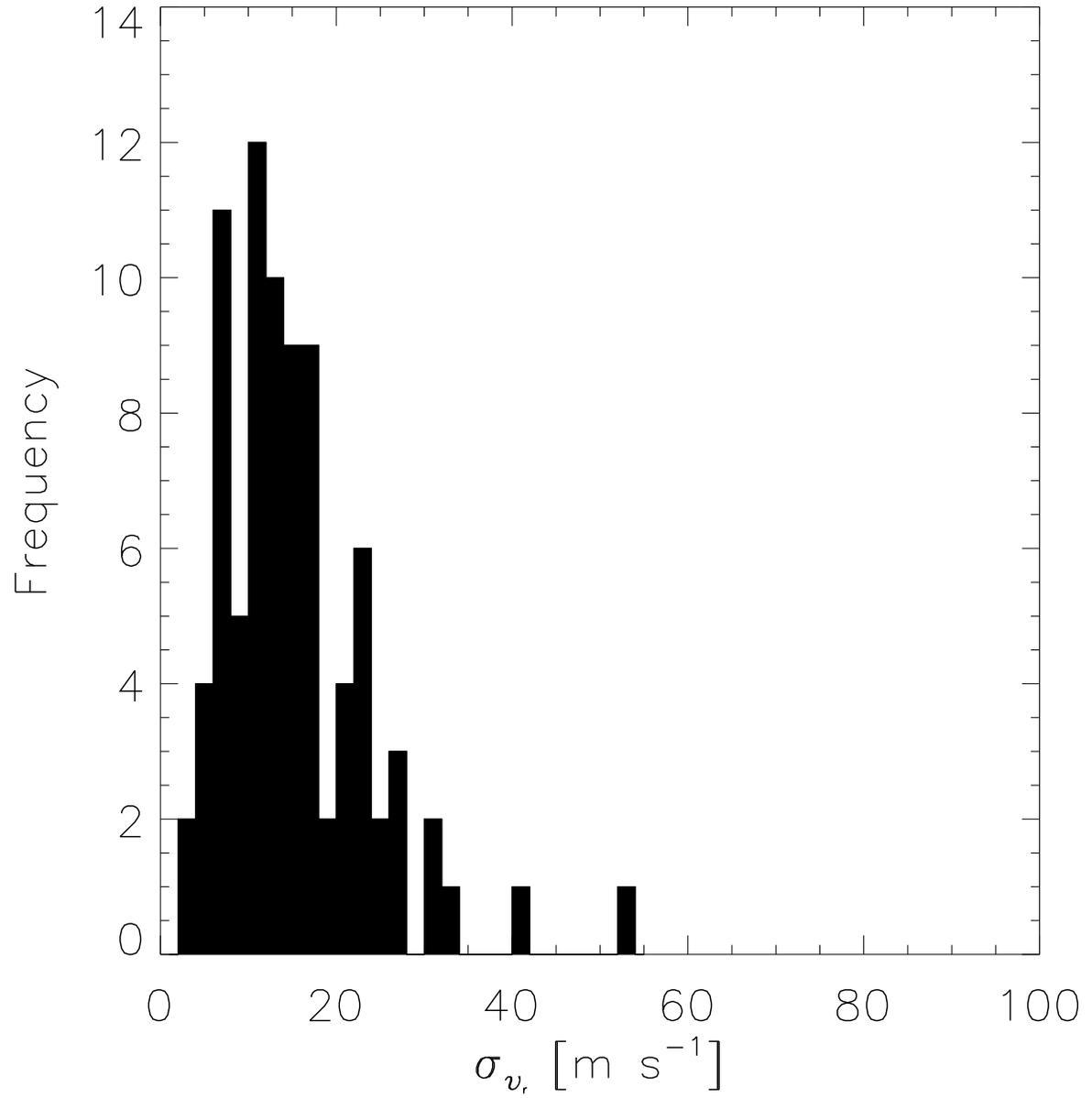}
\caption[Histogram of the rms of Program Stars]{Histogram of the rms scatter
in the program stars, excluding
binary stars and stars with linear trends.}
\end{figure}
 
\begin{figure}
\plotone{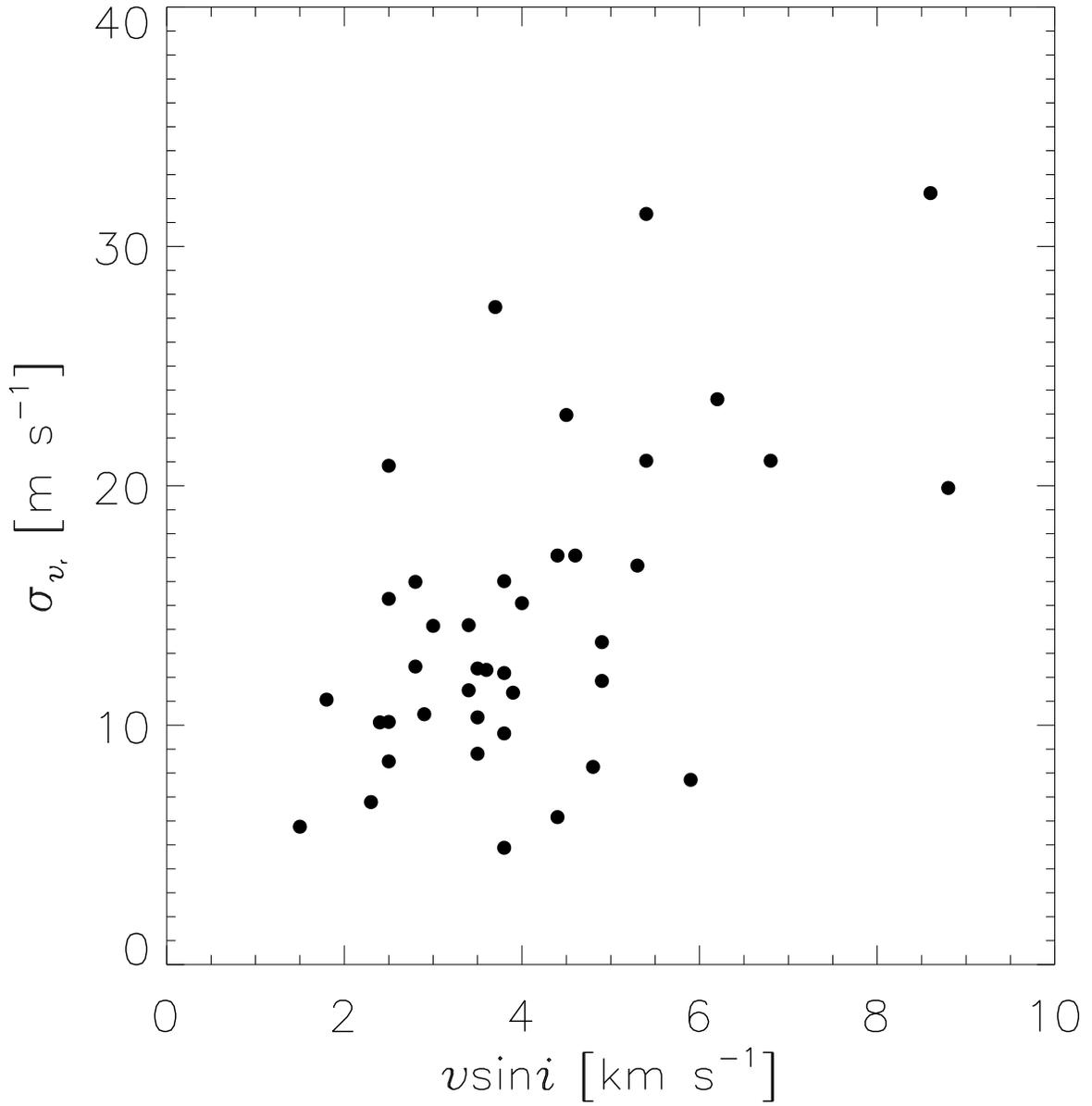}
\caption[rms Scatter vs. $v$sin$i$]{The rms scatter in the program stars
versus the measured
$v$sin$i$. Stars with linear trends and binaries have not been included.}
\end{figure}
 
\begin{figure}
\plotone{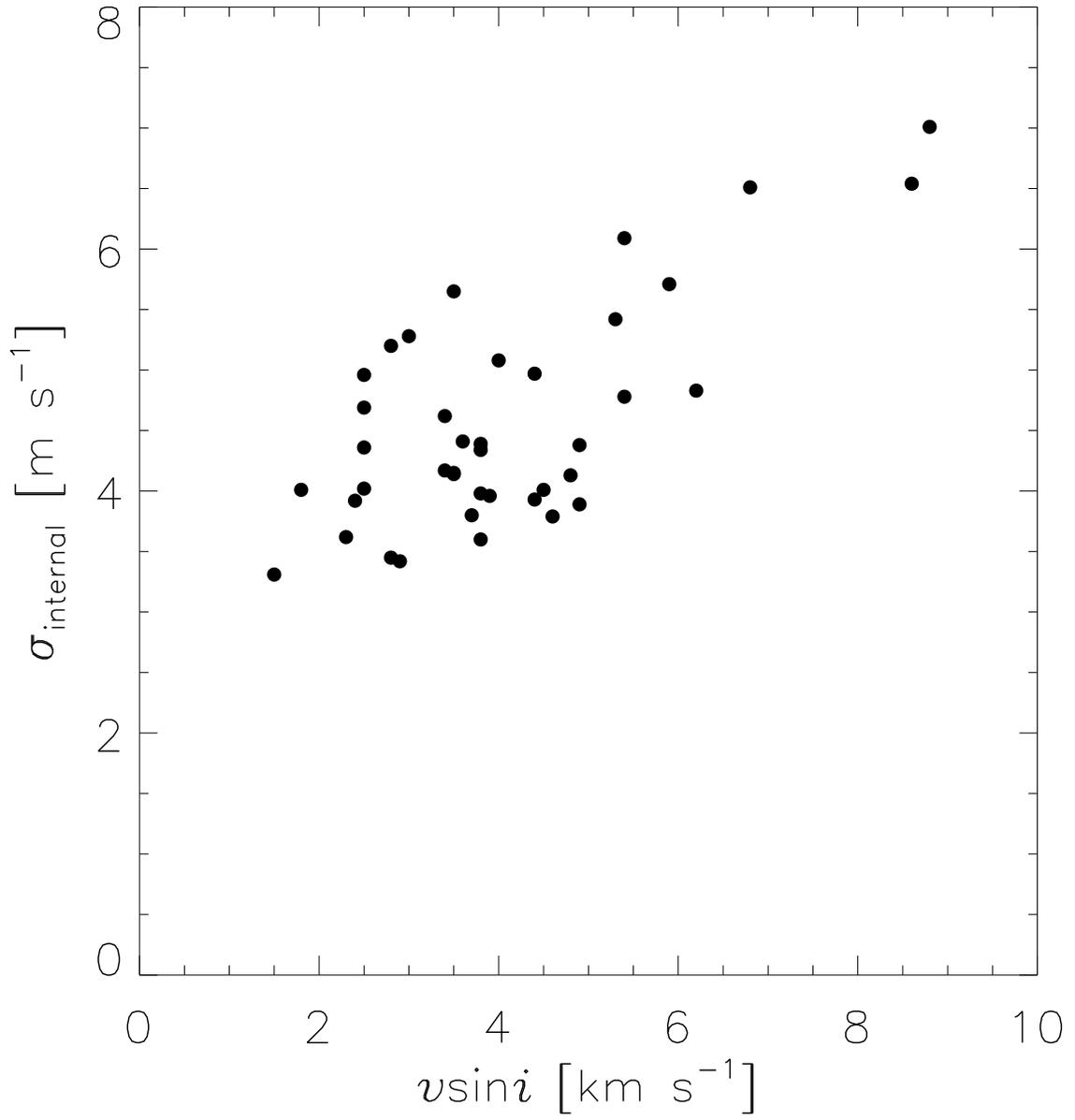}
\caption[Internal errors versus $v$sin$i$]{Internal errors versus $v$sin$i$.}
\end{figure}
 
\begin{figure}
\centering
\scalebox{0.9}{\includegraphics{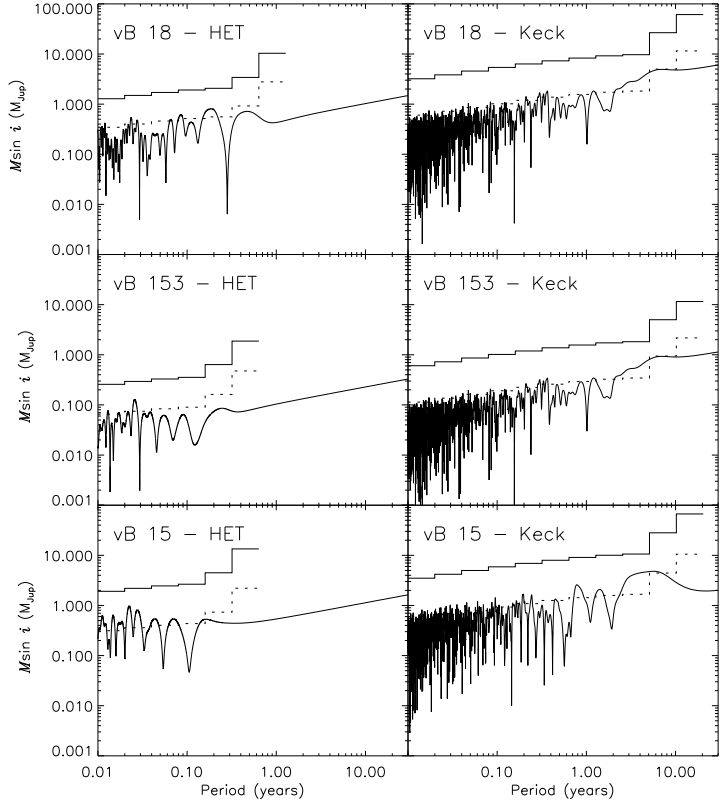}}
\caption[Mass Limits on the Detection of Companions]{The above plots show the
limits of our data on the detection of
companions. The solid curve represents a power spectrum of the data
translated to $M$sin$i$ units. The dashed histogram plot, as discussed in the
text, is a 99\% confidence level for the detection of significant peaks in the
data (where a data peak crossing this line would have a 99\% confidence of
being a true companion). This assumes that the only error for each data point
is the internal
error of the observations (ranging from about 5-7 m s$^{-1}$). The solid
histrogram plot represents a 99\% confidence level considering an average
error as determined from the rms of $v_{\rm r}$ caused by stellar activity.}
\end{figure}

\begin{figure}
\scalebox{0.9}{\includegraphics{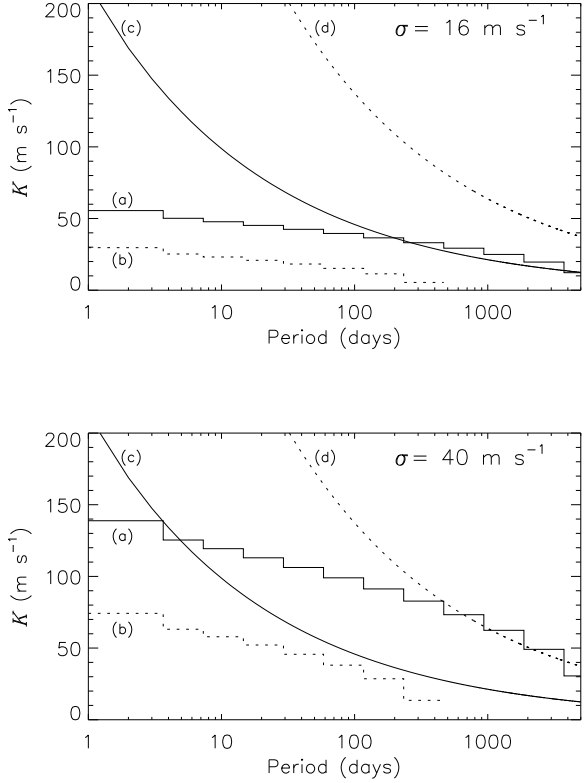}}
\caption{Detectability of companions around Hyades-aged stars. The 2 panels 
represent errors of 16 and 40~m~s$^{-1}$ (inclusive of internal errors and 
radial velocity jitter from stellar activity). Within each panel, the two
histograms indicate sampling similar to our Keck data (a) and 
our HET data (b). For reference, the velocity induced by a 1 M$_{Jup}$ planet 
(c) and a 3 M$_{Jup}$ (d) planet around a 1 M$_{\odot}$ star (in a
circular orbit) for various orbital periods are shown.
As in Figure 5, velocity space above the histograms are detectable. So, for 
example, a 1 M$_{Jup}$ planet is just barely detectable if it has a period of
only a few days when the errors are the highest in our sample (40~m~s$^{-1}$)
but is easily detectable if the same planet has a period of less than 
3 months and the error is ``average" for the Hyades (16~m~s$^{-1}$).}
\end{figure}

\begin{deluxetable}{llrr}
\tabletypesize{\scriptsize}
\tablecaption{Radial Velocity Observations\label{tbl-1}}
\tablewidth{0pt}
\tablehead{
\colhead{Star} & \colhead{JD - 2400000} & \colhead{$v_{\rm r}$} &
\colhead{$\sigma_{v_{\rm r}}$} \\
&days&[m s$^{-1}$] & [m s$^{-1}$]
} 
\startdata
BD+04 810  & 50792.118 & -15.06 &  5.55 \\
 & 51076.098 & -4.51 &  2.82 \\
 & 51441.129 & 5.76 &  4.27 \\
 & 51549.832 & 10.87 &  4.48 \\
 & 51880.897 & -5.16 &  4.05 \\
\enddata
\tablecomments{The complete version of this table is in the electronic
edition of the Journal.  The printed edition contains only a sample.}
\end{deluxetable}

\clearpage
\begin{deluxetable}{lcc}
\tabletypesize{\scriptsize}
\tablecaption{Linear Trends and Binaries\label{tbl-2}}
\tablewidth{0pt}
\tablehead{
\colhead{Star} & \colhead{slope (m s$^{-1}$/JD)} & \colhead{Patience et al.} }
\startdata
vA 486 & 1.7500 & N\\
vB 5 & -0.1364 & Y\\
vB 17 & -0.0909 & Y\\
vB 52 & -0.0333 & Y\\
vB 88 & see Table 3 & N\\
vB 184 & 0.3818 & N\\
\enddata
\end{deluxetable}

\clearpage
\begin{deluxetable}{ll}
\tabletypesize{\scriptsize}
\tablecaption{Orbital Parameters for vB 88\label{tbl-3}}
\tablewidth{0pt}
\tablehead{
\colhead{Parameter} & \colhead{Value}}
\startdata
$m$sin$i$ (M$_{\odot}$) &  0.069\\
Period (days) & 2809.2 $\pm$ 80.7\\
V$_{0}$ (m s$^{-1}$) & -485.3 $\pm$ 5.8\\
T$_{0}$ (JD) & 2452100.01 $\pm$ 18.9 \\
$e$& 0.5166 $\pm$ 0.0123  \\
$\omega$ (degrees)& 136.45 $\pm$ 3.12 \\
K$_{1}$ (m s$^{-1}$) & 1152.2 $\pm $15.0\\
\enddata
\end{deluxetable}
 
\clearpage
\begin{deluxetable}{lccc}
\tabletypesize{\scriptsize}
\tablecaption{$v_{\rm r}$ Data\label{tbl-4}}
\tablewidth{0pt}
\tablehead{
\colhead{Star} & \colhead{$\sigma_{v_{\rm r, int}}$ [m s$^{-1}$]} &
\colhead{$\sigma_{\rm int}$ [m s$^{-1}$]}\\
}
\startdata
BD+04 810   &10.12     &3.92\\
BD+07 499   &15.15     &4.52\\
BD+08 642   &12.94     &4.73\\
BD+17 455   &11.07     &4.01\\
BD+17 719c  &737.05    &4.32\\
BD+19 650   &17.56     &3.47\\
HD 18632    &10.46     &3.42\\
HD 19902    &5.76      &3.31\\
HD 23453    &6.82      &4.42\\
HD 26257    &7.72      &5.71\\
HD 35768    &6.16      &4.97\\
HD 240648   &11.85     &4.38\\
HD 242780   &8.26      &4.13\\
HD 283869   &5.56      &3.65\\
HD 284552   &25.63     &4.60\\
HD 284653   &14.27     &3.62\\
HD 284930   &14.86     &4.36\\
HD 285367   &27.47     &3.80\\
HD 285482   &13.13     &3.73\\
HD 285590   &6.03      &3.98\\
HD 285625   &72.92     &4.91\\
HD 285837   &16.11     &5.92\\
HD 285849   &20.22     &6.54\\
HD 286363   &15.98     &4.38\\
HD 286554   &13.82     &5.18\\
HD 286734   &6.87      &3.37\\
HD 286789   &7.45      &3.83\\
HD 286929   &6.14      &3.99\\
HIP 15720   &5.04      &4.66\\
HIP 16548   &17.31     &7.11\\
HIP 17766   &9.66      &3.71\\
HIP 19082   &22.57     &4.98\\
HIP 22177   &9.66      &5.71\\
J 303       &3.10      &4.79\\
J 332       &12.08     &4.75\\
J 348       &7.40      &4.54\\
vA 115      &22.81     &7.39\\
vA 146      &10.45     &5.40\\
vA 383      &5.31      &4.64\\
vA 486\tablenotemark{1}&231.93&39.20\\
vA 502      &10.73     &7.19\\
vA 529      &5.91      &9.47\\
vA 637      &22.14     &6.96\\
vA 638      &8.98      &5.84\\
vA 731      &2.63      &4.99\\
vB 1        &29.87     &4.55\\
vB 2        &19.03     &4.69\\
vB 4        &13.47     &3.45\\
vB 5\tablenotemark{1} &25.26&3.85\\
vB 7        &16.02     &3.60\\
vB 10       &23.62     &4.83\\
vB 12       &20.84     &4.02\\
vB 15       &40.39     &5.12\\
vB 17\tablenotemark{1}&21.44&4.99\\
vB 18       &31.36     &4.78\\
vB 19       &17.11     &11.49\\
vB 21       &6.79      &3.62\\
vB 25       &8.49      &4.36\\
vB 26       &10.33     &4.15\\
vB 27       &15.99     &3.89\\
vB 31       &27.46     &6.54\\
vB 42       &12.45     &4.34\\
vB 46       &4.88      &5.28\\
vB 48       &21.76     &8.92\\
vB 49       &14.15     &5.20\\
vB 52\tablenotemark{1} &15.99&5.81\\
vB 65       &19.91     &7.01\\
vB 66       &32.23     &6.54\\
vB 73       &21.05     &6.51\\
vB 76       &10.14     &4.69\\
vB 79       &14.87     &4.73\\
vB 87       &15.10     &5.18\\
vB 88\tablenotemark{2}&18.44&9.67\\
vB 92       &9.66      &4.39\\
vB 93       &53.15     &6.09\\
vB 97       &21.05     &6.09\\
vB 99       &11.46     &4.17\\
vB 105      &15.28     &4.96\\
vB 109      &17.09     &3.79\\
vB 118      &16.67     &5.42\\
vB 127      &12.18     &3.98\\
vB 143      &13.92     &7.76\\
vB 153      &22.96     &4.01\\
vB 170      &17.24     &4.09\\
vB 173      &23.99     &3.85\\
vB 174      &12.02     &3.75\\
vB 178      &12.37     &4.14\\
vB 179      &12.31     &4.41\\
vB 180      &11.36     &3.96\\
vB 183      &8.81      &5.65\\
vB 184\tablenotemark{1}&40.52&6.16\\
vB 187      &17.09     &3.93\\
vB 191      &2.91      &3.93\\
\enddata
\tablenotetext{1}{$\sigma_{v_{\rm r}}$ and $\sigma_{\rm int}$ listed are
residuals of the stellar data with linear trends (slopes given in Table 2) 
removed.}
\tablenotetext{2}{$\sigma_{v_{\rm r}}$ and $\sigma_{\rm int}$ listed are
for residuals of the data with the orbital parameters listed in Table 3 
removed.}
\end{deluxetable}
 
\clearpage
\begin{deluxetable}{lccc}
\tabletypesize{\scriptsize}
\tablecaption{Stars with FAP $<$10\% \label{tbl-5}}
\tablewidth{0pt}
\tablehead{
\colhead{Star} & \colhead{P (days)} & \colhead{FAP$_{H\&B}$} &
\colhead{FAP$_{bootstrap}$}}
\startdata
vB 7 &9.61&0.088&0.045\\
 vB 12 &4.04&0.097&0.001\\
 vB 18 &17.36&0.040&0.089\\
 vB 19 &6.02&0.056&0.006\\
 &4.91&0.089&0.017\\
vB 87 &7.60&0.069&0.088\\
 vB 118 &6.15&0.068&0.037\\
 vB 153 &4.62&0.010&0.011\\
 vB 170 &26.24&0.053&0.014\\
\enddata
\end{deluxetable}

\clearpage
\begin{deluxetable}{lcccccc}
\tabletypesize{\scriptsize}
\tablecaption{Periods Derived vs. Periods in Literature\label{tbl-6}}
\tablewidth{0pt}
\tablehead{
\colhead{Star} & \colhead{$P_{\rm ours, Keck}$ (d)} &
\colhead{$P_{\rm HET}$ (d)} & \colhead{$P_{\rm lit}$ (d)} &
\colhead{Reference}& \colhead{$P_{\rm pred}$ (d)} & \colhead{Notes}}
\startdata
vB 21 & 5.49&  & 9               &1      &       &       \\
vB 25 & 4.91&  & 12.6            &1      &       &       \\
vB 26 & 4.6&   & 9.3, 9.4, 9.1   &2,1,1  &11.22  &alias?\\
vB 31 & 4.72&  & 5.4             &1      &4.6    &       \\
vB 52 & 5.64&  & 7.9, 8.0        &2,3  &5.04   &       \\
vB 65 & 4.65&  & 5.9             &1      &5.51   &       \\
vB 73 & 12.92& & 7.4             &2,3 &5.24   &       \\
vB 79 & 6.13&  & 11.4, 9.7       &2,3  &12.56  &alias?\\
vB 92 & 22.53& & 9               &1      &9.9    &       \\
vB 97 & 6.45&  & 8.5             &2,3  &7.46   &\\ 
vB 173 &20.82& & 14.1            &1      &       &       \\
vB 174 &10.12& & 11.9            &1      &       &       \\ \hline
vB 15 & 7.43& 8.18 &             &       &6.22   &        \\
vB 18&  17.36& 8.65&             &       &       & alias?\\
vB 153& 4.63&  9.42&             &        &       & alias?\\
\enddata
\tablenotetext{1}{\citet{RaThLo87}}
\tablenotetext{2}{\citet{LoThRa84}}
\tablenotetext{3}{\citet{RaLoSk95}}
\end{deluxetable}
 
\end{document}